%Paper: hep-th/9307013
%From: udah130@elm.cc.kcl.ac.uk
%Date: Fri, 02 Jul 1993 11:58:00 EDT

\input manumac.tex

\twelvepoint

\def\dplus{=\hskip-5pt \raise 0.7pt\hbox{${}_\vert$} ^{\phantom 7}}
\def\dplusup{=\hskip-5.1pt \raise 5.4pt\hbox{${}_\vert$} ^{\phantom 7}}

%\font\twelvebb=msbm10 scaled 1200
%\font\ninebb=msbm10 scaled 900
%\font\sevenbb=msbm10 scaled 700
%\newfam\bbfam
%\def\bb{\fam\bbfam\twelvebb}
%\textfont\bbfam=\twelvebb
%\scriptfont\bbfam=\ninebb
%\scriptscriptfont\bbfam=\sevenbb

\def\notequiv{\equiv\hskip-11pt\slash^{\phantom 8}}

\def\xis{{X}}

%%%the following are the defs from the p,q paper

%\def\pplevel{ {\hskip-4.5pt\vrule height6pt width1pt depth1pt\hskip4.5pt} }

\def\dpl{D_+}
\def\dmi{D_-}

\def\np{\nabla_+}
\def\nm{\nabla_-}

\def\gij{g_{ij}}

\def\bij{b_{ij}}
\def\ffi{\phi^i}
\def\fj{\phi^j}

\def\intx{\int\!\!d^2x\,}

\def\thp{\theta^+}
\def\thm{\theta^-}

\line{\hfil QMW-93-11}
\line{\hfil DAMTP/R-93/8}
\line{\hfil KCL-93-5}
\line{\hfil May 1993}

\vskip 0.4truecm
\centerline{\bf Potentials for (p,0) and (1,1) supersymmetric sigma models with
torsion} \vskip 1truecm
\centerline{{\bf C.M. Hull$^1$, G. Papadopoulos$^2$ and P.K. Townsend$^3$}}
\bigskip\bigskip
{\it
\leftline{$^1$ Physics Department,
Queen Mary and Westfield College, London, England}
\leftline{$^2$ Dept. of Mathematics, King's College, London,
England}
\leftline{$^3$ DAMTP, University of Cambridge, Silver Street, Cambridge,
England}
}
\vskip 0.8truecm
\vskip 1cm

\AB
Using (1,0) superfield methods, we determine the general scalar potential
consistent with off-shell (p,0) supersymmetry and (1,1) supersymmetry in
two-dimensional non-linear sigma models with torsion. We also present an
extended
superfield formulation of the (p,0) models and show how the (1,1) models can be
obtained from the (1,1)-superspace formulation of the gauged, but massless,
(1,1)
sigma model.

\AE

\vfill\eject
\noindent {\bf 1. Introduction }
\medskip
The bosonic two-dimensional non-linear sigma-model with target space
${\cal M}$ is a field theory with fields $\phi$ that are maps from the
two-dimensional Minkowski spacetime to a Riemannian manifold ${\cal M}$. The
general action with second-order field equations is determined by three tensors
on
${\cal M}$. These are (i) the metric tensor $g$, (ii) a closed three-form
$H$, which is interpreted as a torsion tensor, and (iii) a scalar V.
In light-cone coordinates $x^\dplusup, x^=$ for the two-dimensional spacetime
(where the number of plus or minus signs indicates the $SO(1,1)$ charge) the
action is
$$
S=\int\!\! d^2x\big\{
\partial_\dplus\!\!\phi^i\partial_=\phi^j(g_{ij} +b_{ij}) -V(\phi)\big\}
\eqno (1.1)
$$
where $b_{ij}$ is a locally defined potential for $H$: $H_{ijk}=
(3/2)\partial_{[i}b_{jk]}$ ($i,j,k=1\cdots D$, $D={\rm dim}{\cal M}$). When
$V=0$ the model is classically conformal invariant. Of particular interest,
e.g. for applications to superstring theory, are the (p,q)-supersymmetric
extensions of (1.1), which have $p$  spinor charges of one chirality
and $q$ spinor charges of the other chirality. Two-dimensional supersymmetric
sigma models with $V\ne0$ were first investigated about ten years ago [1] for
$p=q$
and zero torsion as a means of introducing an infrared regulator into the
massless
models. In recent years (2,2) models with $V\ne0$ have been investigated in
connection with Landau-Ginsberg formulations and
integrable deformations of N=2 supersymmetric conformal field theory [2]. Of
particular interest   is the fact that the deformed  models have
solitons
stabilized by a complex topological charge which appears in the supersymmetry
algebra as a central charge. There also exist (4,4)-supersymmetric models
with $V\ne0$ with solitons carrying a quaternionic charge [3].

{\sl All} two-dimensional supersymmetric sigma models, with or without
potential,
are special cases of the general (1,0) model, which can be written in terms
of (1,0) scalar superfields $\{\phi^i(x,\theta^+); i=1,\dots,D\}$ and spinor
superfields $\{\psi_-^a(x,\theta ^+); a=1,\dots,n\}$.
The superfields $\phi^i$ define a map $\phi$ from (1,0) superspace,
$\Sigma^{(1,0)}$, into ${\cal M}$. The superfield $\psi$ is a section of
the vector bundle $S_{-}\otimes \phi^*\xi$ where $\xi$ is a vector bundle over
${\cal M}$ of rank $n$ and $S_-$ is the spin bundle over $\Sigma^{(1,0)}$.
The (1,0)-superspace action without a potential has been given in previous work
[4]. The generalization to include a potential is\footnote{$^\dagger$}{The zero
of the energy can be shifted by an arbitrary constant by the inclusion in the
superspace Lagrangian of a constant proportional to $\theta^+$, but this would
introduce a {\sl Lorentz non-covariant} central charge into the supersymmetry
algebra so we omit it.}
$$
S=\int\! d^2 x d \theta ^+ \big\{ D_+\phi^i\partial_=\phi^j (g_{ij} +b_{ij})
+i\psi_-^a\nabla_+\psi_-^b \; h_{ab} +ims_a\psi_-^a\big\}\ ,
\eqno (1.2)
$$
where  $D_+$ is the spinor derivative satisfying $D_+^2=i\partial_\dplus$,
$h_{ab}$ is a fibre metric of $\xi$ and
$$
\nabla_+\psi_-^b\equiv (D_+\psi_-^b + D_+\phi^i \Omega_{i}{}^b{}_c\psi_-^c)\ ,
\eqno (1.3)
$$
where $\Omega(\phi)$ is a connection for $\xi$. The parameter $m$ is a constant
with dimensions of mass and $s_a$ is a section of $\xi^*$.
Without loss of generality one may choose a connection such that the fibre
metric
is covariantly constant, i.e. $\nabla_i h_{ab}=0$.

The component action corresponding to (1.2) can be obtained by standard
methods.
After elimination of auxiliary fields one finds that the potential is given in
terms of $s$ by
$$
V(\phi) ={1\over4}m^2 h^{ab}s_as_b\ .
\eqno (1.4)
$$
We shall restrict $h_{ab}$ to be positive definite, in which case the potential
is positive semi-definite and the structure group of the bundle $\xi$ is a
subgroup of $O(n)$. The potential vanishes at the zeros of $s_a(\phi)$ which
are
therefore the classical `vacua' of the model. For many models of interest
$s_a$
will have isolated zeros.\footnote{$^\dagger$}{Note that generic sections of
vector
bundles over compact manifolds have isolated zeros.} If there is more than one
isolated zero then there will be soliton solutions interpolating between them,
as
happens for the specific (2,2) and (4,4) models discussed in [3].

In the $m=0$ case a (1,0)-supersymmetric sigma-model will be invariant under
(p,q) supersymmetry provided that the target space satisfies certain
conditions that have a natural interpretation in the language of complex
geometry [5]. Similarly for $m\ne0$, the requirement of (p,q) supersymmetry
will
place restrictions on the couplings $g_{ij}$, $b_{ij}$, $h_{ab}$ and $s_a$. The
purpose of this paper is to find and analyse these conditions for (p,0) and
(1,1)
supersymmetry. Results on some of the (2,0) models that we obtain have already
appeared [6] during the preparation of this paper.

The (p,0) case can be analysed either in (1,0) superspace or by a
straightforward extension of the extended superfield methods of ref.[7],
because
the (p,0) algebra cannot develop a central charge; we shall discuss both
methods.
The inclusion of a scalar potential in models with supersymmetry charges of
{\sl
both} chiralities, of which the (1,1) models are the simplest examples, is
complicated by the fact that to obtain the general scalar potential one must
include the possibility of central charges. The (1,1) model will be analysed in
both (1,0) superspace and (1,1) superspace. Standard (1,1) superfield
techniques
exclude the possibility of central charges and therefore fail to give the {\sl
general} scalar potential, so a new approach is presented here that makes use
of
the (1,1) superfield methods of [8] and the geometry of supersymmetric gauged
sigma models developed in [9].

\medskip
\noindent {\bf 2. (p,0) Supersymmetry}
\medskip

The variation of (1.2) with respect to the {\sl arbitrary}
variations $\delta\phi^i$ and $\delta\psi_-^a$ of $\phi^i$ and $\psi_-^a$ is
(up to a surface term)
$$
\delta S=\int\! d^2 xd\theta^+\big\{ \delta\phi^i {\cal S}_{-i}
+\Delta\psi_-^a{\cal
S}_a\big\}\  ,
\eqno (2.1)
$$
where
$$
\Delta\psi_-^a\equiv \delta\psi_-^a +\delta\phi^i\psi_-^b\Omega_i{}^a{}_b
\eqno (2.2)
$$
is the covariantization of $\delta\psi_-^a$, and
$$
\eqalign{
{\cal S}_{-i} &\equiv -2g_{ij}\nabla^{(-)}_+\partial_=\phi^j -i\psi_-^a\psi_-^b
D_+\phi^j
F_{ijab} + im\nabla_i s_a\psi_-^a \cr
{\cal S}^a &\equiv 2i\nabla_+\psi_-^a +ims^a \ .\cr}
\eqno (2.3)
$$
The covariant derivatives involve the connections with torsion given by
$\Gamma^{(\pm)}_{jk}{}^i= \Gamma_{jk}{}^i \pm H^i{}_{jk}$ where $\Gamma$ is the
standard Levi-Civita connection. Using (2.3) it is readily verified that the
action (1.2) is invariant under the transformations
$$
\eqalign{
\delta_\epsilon\phi^i &= -{i\over2}D_+\epsilon_= \; D_+\phi^i
+\epsilon_=\partial_\dplus\!\phi^i\cr
\Delta_\epsilon\psi_-^a &=-{i\over2} D_+\epsilon_=\;
\nabla_+\psi_-^a +\epsilon_=\nabla_\dplus\!\psi_-^a\cr}
\eqno (2.4)
$$
for $x$-independent (but $\theta$-dependent) superfield parameter $\epsilon_=$.
These transformations are those of (1,0) supersymmetry together with
translations
in the $x^\dplusup$ direction.

Any additional supersymmetries of (1.2) of the same chirality must have Noether
charges that anticommute with the first one. This implies that the additional
supersymmetry transformations can be expressed in terms of (1,0) superfields
and a
set of constant, anticommuting, parameter(s) $\{\eta^r_- \ ;r=1,\dots ,p-1\}$.
The form of these transformations for $m=0$ is fixed by dimensional
analysis; when $m\ne 0$ we must allow for an additional fermion variation
proportional to $m$. We are thus led to consider
$$
\eqalign{
\delta_\eta\phi^i &=i\eta^r_- I_{r}{}^i{}_j(\phi)D_+\phi^j\cr
\Delta_\eta\psi_-^a &={1\over2}\eta^r_-{\hat I}_r{}^a{}_b(\phi){\cal S}^b
+{im\over2}\eta^r_- G_r^a(\phi)\cr}
\eqno (2.5)
$$
where $I_r$ are tensors on ${\cal M}$, and $G_r$ and $\hat I_r$  are
sections of $\xi^*$ (the dual of $\xi$) and $\xi\otimes \xi^*$, respectively.
The field equation term in the above transformations is not necessary for a
determination of the conditions arising from {\sl on-shell} closure of the
(p,0)
algebra (see, for example, the discussion of the massless (2,0) models in ref.
[4]). However, this term {\sl is} required for off-shell closure [7], and hence
for an
extended superspace formulation to be possible.

The conditions required for off-shell closure [7] on $\phi$ are
$$
I_rI_s = -\delta_{rs} + f_{rs}{}^t I_t
\eqno (2.6)
$$
(in matrix notation) and
$$
N(I_r,I_s)^i{}_{jk} =0\ ,
\eqno (2.7)
$$
where $f_{rs}{}^t$ is zero for p=2 and equal to the quaternion structure
constants
$\epsilon_{rst}$ for p=4, and $N$ is the
generalised Nijenhuis tensor defined by
$$
N(I_r,I_s)^i{}_{jk} \equiv 2\big[\partial_l I_r{}^i{}_{[k} I_s{}^l{}_{j]}
- I_r{}^i{}_l \partial_{[j} I_s{}^l{}_{k]} + (r\leftrightarrow s)\big]\ .
\eqno (2.8)
$$
Off-shell closure on $\psi_-^a$ requires
$$
F_{ij}{}^a{}_b I_r{}^i{}_{[k} I_s{}^j{}_{l ]} = F_{kl}{}^a{}_b\; \delta_{rs}
\eqno (2.9)
$$
and
$$
\hat I_r{}^a{}_c\hat I_s{}^c{}_b =-\delta_{rs} \delta^a_b+
f_{rs}{}^t \hat I_t{}^a{}_b
\eqno (2.10)
$$
and
$$
I_r{}^j{}_i\nabla_j\hat I_s{}^a{}_b -
\hat I_r{}^a{}_c\nabla_i\hat I_s{}^c{}_b + (r \leftrightarrow s) =0\ .
\eqno (2.11)
$$
The above (off-shell) conditions are sufficient when $m=0$ [7]. When $m\ne0$ we
find the additional condition
$$
\big[I_r{}^i{}_j \nabla_i G_s^a + (r\leftrightarrow s)\big]
+2\delta_{rs} \nabla_i s^a =0\ .
\eqno (2.12)
$$

When $m=0$ the conditions for invariance of the action are
$$
I_r{}^k{}_{(i}g_{j)k} =0 \qquad \nabla_i^{(+)} I_r{}^j{}_k =0
\eqno (2.13)
$$
and
$$
\hat I_r{}_{(ab)}\equiv h_{c(a}\hat I_r{}^c{}_{b)} =0\ .
\eqno (2.14)
$$
When $m\ne 0$ we require additionally that
$$
\big( G_r^a s_a \big) = {\rm constant}
\eqno (2.15)
$$
and
$$
\nabla_i G_r^a = I_r{}^j{}_i \nabla_j s^a \quad (r=1\dots p-1)\ .
\eqno (2.16)
$$
The integrability condition for (2.16) is (2.9). The condition (2.15) was
stated
previously for the (2,0) model but with the constant set equal to zero [6].

We now turn to the analysis of these conditions. We shall first recall the
(previously established) results for m=0. For p=2 ${\cal M}$ is a complex
manifold
with complex structure $I$; $g$ is an hermitian metric with respect to $I$, and
the
holonomy of the connection $\Gamma^{(+)}$ is a subgroup of $U(D/2)$.
Furthermore,
as implied by (2.9)-(2.11) and (2.14), the vector bundle $\xi$ is holomorphic
and
$\xi$ is hermitian.  For $p=4$ ${\cal M}$ admits a quaternionic structure, i.e
the
three (integrable) complex structures obey the algebra of imaginary unit
quaternions, the metric $g$ is tri-Hermitian and the holonomy of the connection
$\Gamma^{(+)}$ is a subgroup of $Sp(D/4)$. Furthermore, the bundle $\xi$ is
tri-holomorphic and tri-hermitian, i.e. holomorphic and hermitian with respect
to
all three complex structures.

The additional conditions that arise for $m\ne0$ are just (2.12), (2.15) and
(2.16), but (2.12) is implied by (2.6) and (2.16) which leaves (2.15) and
(2.16). The analysis of these two conditions is facilitated by the definitions
$$
\eqalign{
L_r^a &=G_r^a +\hat I_r{}^a{}_b s^b\cr
M_r^a &=G_r^a -\hat I_r{}^a{}_b s^b}
\eqno (2.17)
$$
in terms of which we have
$$
\eqalign{
G_r^a &={1\over2}\big(M_r^b +L_r^b\big)\cr
s^a &={1\over2}\hat I_r{}^a{}_b\big(M_r^b -L_r^b\big)}
\eqno (2.18)
$$
where there is {\sl no summation} over $r$ in the second of these equations
(i.e.
for p=4 we have three different expressions for $s^a$). The conditions (2.16)
and (2.15) can now be rewritten in terms of $L_r$ and $M_r$ as follows:
$$
\eqalign{
I_r{}^j{}_i\nabla_j L_r^a &- \hat I_r{}^a{}_b\nabla_i L_r^b =0\cr
I_r{}^j{}_i\nabla_j M_r^a &+ \nabla_i\big( \hat I_r{}^a{}_b M_r^b\big) =
-\big(\nabla_i\hat I_r{}^a{}_b\big) L^b_r}
\eqno (2.19)
$$
and
$$
(\hat I_r)_{ab}L_r^a M_r^b = {\rm constant}\quad (r=1\dots p-1)
\eqno (2.20)
$$
where there is again no summation over $r$. Similarly, the potential $V$ can be
expressed in terms of $L_r$ and $M_r$ as
$$
V= {1\over 16}m^2 h_{ab}(M_r^a-L_r^a)(M_r^b-L_r^b)\ .
\eqno (2.21)
$$

It remains to understand the conditions (2.19) and (2.20) on $L_r$ and $M_r$.
Let
us assume that $\nabla\hat I_r=0$.\footnote{$^\dagger$}{In the (2,0) case it is
always possible to find a metric connection with this property [7].} To solve
(2.19) we choose complex coordinates adapted to the pair $(I_r,\hat
I_r)$ of complex structures. Then
$$
\nabla_{\bar\mu} L_r^\alpha =0\qquad \nabla_{\mu} M_r^\alpha =0
\eqno (2.22)
$$
where $\mu=1,\cdots , (D/2)$ and $\alpha=1,\cdots, (n/2)$. Thus $L_r$ is a
holomorphic and $M_r$ an antiholomorphic section of $\xi$ with respect to
$(I_r,\hat I_r)$. The potential is expressed in terms of these sections as in
(2.21) (and for p=4 this can be done in three different ways).

An interesting special case of this result is found by setting $M_r=0$. In
this case (2.22) implies that $s$ is a holomorphic section of $\xi$ for p=2 and
a triholomorphic section of $\xi$ (i.e. holomorphic with respect to all three
pairs of complex structures) for  p=4. The zeros of the potential (the
`classical vacua') are then given by the zeros of a (tri)holomorphic section
of $\xi$.

Finally, we remark that a more general class of (p,0) models is obtained when
the condition of off-shell closure of the supersymmetry algebra is relaxed to
on-shell closure. Note that the first term in $\Delta_\eta\psi_-^a$ of (2.5) is
trivially a symmetry by itself provided (2.14) is satisfied; it was included as
part of the definition of the supersymmetry transformation in order to ensure
off-shell closure. For on-shell closure it can be dropped, in which case the
conditions (2.10) and (2.11) clearly do not arise and can also be dropped. Then
$\hat I$ is restricted only by (2.14) which indeed allows the choice $\hat
I=0$ made in [4]. The implications of the remaining conditions in this case
will
be discussed elsewhere.

\bigskip
{\bf 3. (p,0) superfields}
\medskip

We shall now give an off-shell (p,0) superspace formulation of the (p,0) models
just described, following the treatment of the massless case in [7]. The (p,0)
superspace $\Sigma^{(p,0)}$ has coordinates $\{x^{\dplus},x^=,\theta^+_l;\;
l=0,r;\; r=1,\dots,p-1\}$ and the supercovariant derivatives $D_{l+}$ satisfy
$$
\{D_{l+},D_{m+}\}=2 i \partial_\dplus \delta_{lm}\ .
\eqno (3.1)
$$
The (off-shell) (p,0) superfields $\phi$ and $\psi$ are defined as follows:
$\phi$ is a map
from  $\Sigma^{(p,0)}$ into the sigma model manifold ${\cal M}$ and $\psi$ is a
section of
the vector bundle $\phi^*\xi\otimes S_-$ over $\Sigma^{(p,0)}$, where $\xi$ is
a vector bundle
over ${\cal M}$. These superfields satisfy the chirality constraints
$$
\eqalign{
D_{r+}\phi^i &= I_r{}^i{}_j D_{{}_0}\!{}_+\,\phi^j \cr
\nabla_{r+}\psi_-^a &= \hat I_r{}^a{}_b \nabla_{{}_0}\!{}_+\psi_-^b+ {m\over 2}
L^a_r\cr}
\eqno (3.2)
$$
where $I_r$, $\hat I_r$ and $L_r$ are the same tensors as those denoted
previously
by these symbols. The conditions found in the previous section for off-shell
closure of the (p,0)-supersymmetry transformations can now be interpreted as
the integrability conditions of the above constraints.

In terms of these (p,0) superfields, the massive (p,0)-supersymmetric action is
$$
S=\int\! d^2 x d\theta_{{}_0}^+ \big\{ D_{{}_0}\!{}_+\phi^i\partial_=\phi^j
(g_{ij} +b_{ij})
+i\psi_-^a\nabla_{{}_0}\!{}_+\psi_-^b \, h_{ab} +ims_a\psi_-^a\big\}\ .
\eqno (3.3)
$$
This is similar to the action (1.2) for the (1,0) supersymmetric
sigma models, but note that for $p>1$ the superspace measure appearing in
(3.3) is not the full superspace one. Nevertheless, the action is invariant
under
the full (p,0) supersymmetry transformations provided that the couplings
satisfy
all the conditions previously obtained.

\bigskip
\noindent {\bf 4. (1,1) Supersymmetry } \medskip

Our task now is to find the conditions under which the (1,0)-supersymmetric
action (1.2) has an additional supersymmetry of the {\it opposite} chirality.
In
this case we must also allow for a central charge which
generates an isometry symmetry. On dimensional grounds the
(1,0) superfield form of these additional transformations must have the form
$$
\eqalign{
\delta_\zeta\phi^i &=D_+\zeta e^i{}_a(\phi) \psi_-^a +m\zeta \xis^i(\phi)\cr
\Delta_\zeta\psi_-^a &=-iD_+\zeta e^a{}_i(\phi)\partial_=\phi^i +D_+\zeta
M^a{}_{bc}(\phi)
\psi_-^b\psi_-^c +m\zeta U^a{}_b \psi_-^b\cr}
\eqno (4.1)
$$
where $e_i{}^a$, $e^a{}_i$, $M^a{}_{bc}$, $U^a{}_b$ and $\xis^i$ are globally
defined tensors on ${\cal M}$ and/or the bundle $\xi$. The parameter $\zeta$ is
an
$x$-independent (1,0) Grassman even scalar superfield with expansion
$\zeta= \alpha + \theta ^+ \epsilon_+ $, so that it combines the left-handed
supersymmetry parameter $\epsilon_+$ and the parameter $ \alpha$ of
central charge transformations.

We begin by checking the commutator of two $\zeta$-transformations.
 For $m=0$ we find that the
following
conditions are required:
$$
e_a{}^i e_j{}^a=\delta^i_j \qquad e_i{}^a e_b{}^i =\delta^a_b
\eqno (4.2)
$$
and
$$
M^a{}_{bc} = e_{[b}{}^i e_{c]}{}^j \big(\partial_ie_j{}^a +\Omega_i{}^a{}_b
e_j{}^b\big) \ .
\eqno (4.3)
$$
When $m\ne0$ we find additionally that
$$
U^a{}_b = e_b{}^i\nabla_i \xis^a - 2\xis^c M^a{}_{bc}\ ,
\eqno (4.4)
$$
where $\xis^a\equiv \xis^i e_i{}^a$. The conditions (4.2) imply that the vector
bundle $\xi$ of the fermionic sector is isomorphic to the tangent bundle of the
target manifold ${\cal M}$, and $e_i{}^a$ can therefore be interpreted as a
vielbein. Thus, the indices of $\xi$ and the tangent bundle of ${\cal M}$
become
interchangeable.

A calculation of the commutator of a (1,0)-supersymmetry transformation with a
$\zeta$-transformation  yields the following result
(using no geometric constraints or
  field equations):
$$
\eqalign{
[\delta_\epsilon,\delta_\zeta]\phi^i &= \big(-{i\over2}D_+\zeta
D_+\epsilon_=\big) m\xis^i\cr
[\delta_\epsilon,\delta_\zeta]\psi_-^a &= \big(-{i\over2}D_+\zeta
D_+\epsilon_=\big)m
U^a{}_b\psi_-^b
-([\delta_\epsilon,\delta_\zeta]\phi^k)\Omega_k{}^a{}_b\psi_-^b\ .\cr}
\eqno (4.5)
$$
The right hand side of these commutators is just a $\zeta$-transformation with
a $\theta$-{\sl independent} parameter, and it is readily verified that such
transformations commute with both (1,0) and $(0,1)$ supersymmetries, i.e. {\sl
the
supersymmetry algebra has an off-shell central charge} when both $m\ne0$ and
$\xis^i\notequiv 0$.

We now turn to the conditions obtained from $\zeta$-invariance of the action.
It is very convenient to simplify them with the aid of the conditions already
found
above from closure of the algebra. Doing this one finds that some of them are
not
independent of those already derived above. For $m=0$ the independent ones can
most
easily be summarised by the following equations
$$
h_{ab} = e_a{}^i e_b{}^j g_{ij}
\eqno (4.6)
$$
and
$$
\nabla^{(-)}_j e_i{}^a=0\ .
\eqno (4.7)
$$
When $m\ne0$ we find in addition that the Lie derivatives of both $g_{ij}$ and
$H_{ijk}$ with respect to $\xis^i$ must vanish, so that
$$
\nabla_{(i} \xis_{j)} =0\ ,
\eqno (4.8)
$$
and $\iota_\xis H$ is a closed two-form which implies that there is some
locally
defined one-form $u$ on ${\cal M}$ such that $\iota_\xis H = du$, i.e.
$$
\xis^k H_{ijk} =\partial_{[i}u_{j]}  \ .
\eqno (4.9)
$$
The one-form $u$ is defined by this equation up to the derivative of a local
function. Furthermore,
$$
\xis^i\nabla_i^{(-)} s_j + s^i\nabla_{[i}^{(-)}s_{j]} =0\ ,
\eqno (4.10)
$$
and
$$
s_i =   u_i - \xis_i\ .
\eqno (4.11)
$$

The solution of the entire set of equations resulting from closure of the
algebra and invariance of the action can be stated as follows. Firstly, for
$m=0$,
the quantities $e_i{}^a$ and $e_a{}^i$ are vielbeins that relate the metric $g$
of
${\cal M}$ with the fibre metric $h$ of the bundle $\xi$. The connection
$\Omega$ is
the spin-connection of $\Gamma^{(-)}$, and $M_{ijk}= -H_{ijk}$. For $m\ne0$ we
also
have that the vector $\xis$ is a Killing vector field, that the torsion
$H_{ijk}$ is
invariant with respect to the symmetry generated by this Killing vector field,
that
the section $s_i$ is given by (4.11), and that, after some computation,
$$
U_{ij}= \nabla^{(+)}_{[i}\xis_{j]}=\nabla_{[i} u_{j]} -\nabla_{[i} \xis_{j]}\ .
\eqno (4.12)
$$
{\sl In addition} eq. (4.10) is equivalent (after more computation) to
$$
\partial_i(\xis^j u_j)=0\ .
\eqno (4.13)
$$
Contracting (4.9) with the Killing vector field  $\xis$ and using (4.13) we
learn
that the one-form $u$ is invariant with respect to the symmetry generated by
$\xis$, i.e.
$$
{\cal L}_{{}_\xis} u=0\ .
\eqno (4.14)
$$

{}From eq. (1.4) and using (4.11) we find that
$$
V(\phi) = {1\over4} m^2  g^{ij} (u-\xis)_i(u-\xis)_j\ .
\eqno (4.15)
$$
Using (4.13) we can rewrite the potential as
$$
V= {1\over4} m^2  \big(g^{ij}u_iu_j  + g_{ij}\xis^i\xis^j \big) +\ {\rm
const.}\ .
\eqno (4.16)
$$
where the constant is $-(m^2/2)u\cdot X$.
Note also that because of (4.14), the $u^2$ part of the potential, and hence
the
entire potential, is required to be invariant under the symmetry generated by
$\xis$. In the special case for which the torsion vanishes, and hence for which
$u_i=\partial_i U$ for locally-defined scalar $U$, we have
$$
V(\phi) = {1\over4} m^2 \big( g^{ij}\partial_i U\partial_j U +
g_{ij}\xis^i\xis^j\big)  +\ {\rm const.} \ ,
\eqno (4.17)
$$
which agrees with [1].

\bigskip
\noindent {\bf 5. Potentials from gauged (1,1) sigma models}
\medskip

The (1,1) supersymmetry algebra with central charges is
$$
\{\tilde Q_+ , \tilde Q_+ \}=2 P_\dplus \qquad
\{\tilde Q_- ,\tilde Q_- \}=2 P_=
\qquad \{\tilde Q_+,\tilde Q_-\}= Z \ ,
\eqno (5.1)
$$
and these supercharges can be realised in (1,1)-superspace as
$$
\eqalign{
\tilde  Q_+ &=   Q_+ + {1 \over 2} \theta^- Z
\cr
\tilde Q_- &=   Q_- + {1 \over 2} \theta^+ Z}
\eqno (5.2)
$$
where
$$\eqalign{
  Q_+ &= { \partial \over \partial \theta^+ }
+
i \theta^+ { \partial \over \partial x^\dplusup }
\cr
  Q_- &= { \partial \over \partial \theta^- }
+
i \theta^+ { \partial \over \partial x^= } \ .}
\eqno (5.3)
$$
The $Q_\pm$ generate the usual (1,1) supersymmetry transformations
without a central charge and $Z$ is the central charge generator
which acts by a diffeomorphism along a vector field $X$. This vector
field should leave the action invariant and this implies that $X$ is
Killing and satisfies (4.9).

It would clearly be desirable to have a (1,1) superspace formulation of the
(1,1) models just described but the conventional (1,1) superspace formulation
does not allow for central charges. This problem can be circumvented by using
the
superspace with central charges [8] and a relation between
supersymmetric sigma-models with potentials and gauged (1,1)-supersymmetric
sigma-models. Specifically, under certain conditions the central charge
transformations can be thought of as certain gauge transformations.

The action of the gauged (1,1) supersymmetric sigma model with target manifold
${\cal M}$ and gauge group $G$ is given by [9]
$$
\eqalign{
 S = \intx d\thp d\thm\Big(&\gij\np\ffi\nm\fj + \bij\dpl\ffi\dmi\fj\cr
&   -{1\over2} A_+  u_{i} \dmi\ffi  -{1\over2} A_-  u_{i} \dpl\ffi\,
+\, m W(\phi)\Big)}
\eqno (5.4)
$$
where $\phi^i$ are now (1,1) superfields, the couplings $g$ and $b$ are defined
as in the action (2.1),  $u_i$ is defined as in (4.9), and $W$ is a scalar on
${\cal M}$. The superfields $A_\pm$ are the superspace components of the
connection $A$ for $G$ which here we take to be one-dimensional abelian. The
derivatives $\nabla_\pm$ are the associated covariant derivatives. In
particular
$\nabla_\pm \phi^i = D_\pm\phi^i + A_\pm \xis^i$, where $D_\pm$ are the spinor
derivatives of (1,1) superspace ($D_+^2=i\partial_\dplus, D_-^2=i\partial_=$)
and
$\xis$ is the vector field on ${\cal M}$ generated by the action of the gauge
group.

This action is invariant under the gauge transformations
$$
\delta \phi ^i = \lambda \xis^i, \qquad \delta A_\pm = - D_\pm  \lambda
\eqno (5.5)
$$
for arbitrary superfield parameter $ \lambda(x, \theta)$  provided that the
following conditions hold:
$$
\eqalign{
\nabla_{(i}\xis_{j)}&=0 \qquad {\cal L}_\xis H_{ijk}=0\cr
\xis^i u_i &=0\qquad \xis^i\partial_i W =0}
\eqno (5.6)
$$
where $H$ is the torsion three form. Observe that this action is manifestly
(1,1)
supersymmetric.

We now choose  particular values for the gauge fields  as follows:
$$
A_\pm = \bar A_\pm \equiv {m \over 2} \theta_\pm
\eqno (5.7)
$$
so that the field strength $\bar F_{+-}=m$. Under combined (1,1) supersymmetry
and gauge transformations the connections $\bar A_\pm$ transform as
$$
\delta  \bar A_\pm =  {m \over 2}\epsilon_\pm - D_\pm \lambda
\eqno (5.8)
$$
and so vanish if
$$
\lambda =\bar\lambda \equiv {m \over 2}(\theta ^+ \epsilon ^- + \theta ^-
\epsilon
^+)\ .
\eqno (5.9)
$$
Thus the gauge fields (5.7) are invariant under supersymmetry transformations
supplemented by compensating gauge transformations with parameters (5.9), and
these modified supersymmetry transformations are precisely those generated by
the
modified supersymmetry operators (5.2). This means that a theory that is
invariant under the action of these supersymmetry generators with central
charge
is given by setting the gauge fields to the values (5.7) in the gauged
sigma-model
action. Moreover, this action is in fact invariant if the last two conditions
in
(5.6) are relaxed to become respectively $u\cdot X= {\rm constant}$ and
$X^i\partial_i W={\rm constant}$, in which case the result agrees with that of
the
(1,0) superfield calculation.

\vskip 1cm
\noindent{\bf Acknowledgements}: G. Papadopoulos wishes to thank the commission
of
European communities for financial support.

\vfill\eject
\centerline {\bf References}
\vskip 1cm

\item {[1]}
 L. Alvarez-Gaum{\' e} and D.Z. Freedman, Commun. Math. Phys. {\bf 91} (1983)
87.

\item {[2]}
P. Fendley, W. Lerche, S.D. Mathur and N.P. Warner, Phys. Lett. {\bf 243B}
(1990)
257.

\item {[3]}
E.R.C. Abraham and P.K. Townsend, Phys. Lett. {\bf 291B} (1992) 85; Phys.Lett.
{\bf B295} (1992) 2545.

\item {[4]}
C.M. Hull and E. Witten, Phys. Lett. {\bf 160B} (1985) 398.

\item {[5]}
C.M. Hull, in {\it Super Field Theories}, eds. H. Lee and G. Kunstatter
(Plenum,
N.Y. 1986).

\item {[6]}
E. Witten, {\it Phases of N=2 Theories in Two Dimensions}, preprint
IASSNS-HEP-93/3.

\item {[7]}
P.S. Howe and G. Papadopoulos, Class. Quantum Grav. {\bf 5} (1988) 1647;
Nucl. Phys. {\bf B219} (1987) 264.

\item {[8]}
S.J. Gates, Nucl. Phys. {\bf B238} (1984) 349.

\item {[9]}
C.M. Hull and B. Spence, Phys. Lett. {\bf 232B} (1989)  204; C.M. Hull, G.
Papadopoulos and B. Spence, Nucl. Phys. {\bf B363} (1991) 593.

\end